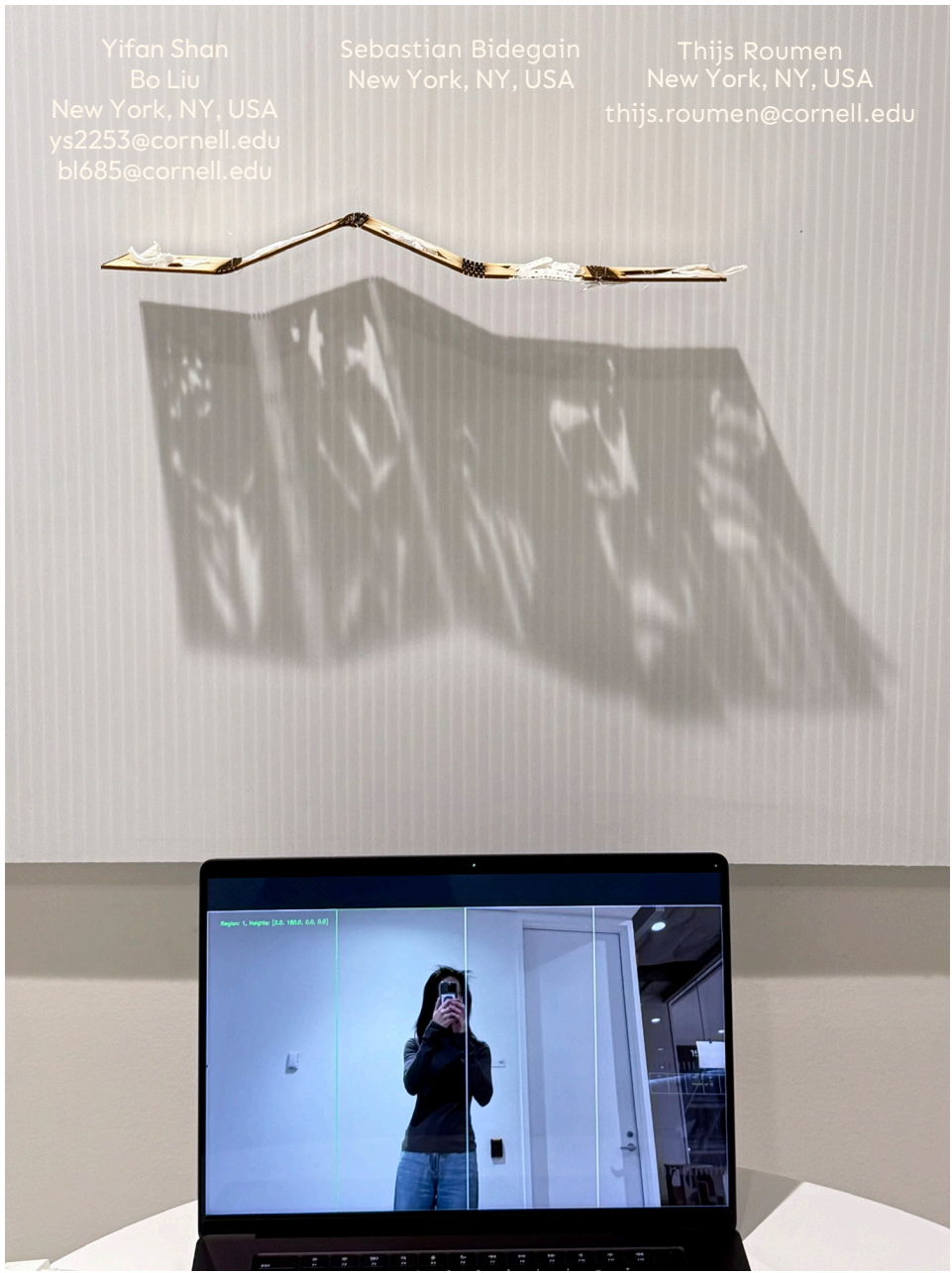

# THE WASTIVE:
## An Interactive Ebb and Flow of Digital Fabrication Waste

Yifan Shan
Bo Liu
New York, NY, USA
ys2253@cornell.edu
bl685@cornell.edu

Sebastian Bidegain
New York, NY, USA

Thijs Roumen
New York, NY, USA
thijs.roumen@cornell.edu

## ABSTRACT
What if digital fabrication waste could observe the world? What would they see? What would they say?

"THE WASTIVE" reimagines digital fabrication waste as sentient observers, giving them a poetic voice through interactive art. As viewers approach, the installation awakens, mimicking the rhythmic ebb and flow of ocean waves — a silent dialogue where discarded materials "observe" and respond to human presence. These interactions echo the gentle murmurs of the sea, transforming technological residue into a reflective, sensory experience.

Through this artistic contemplation, "THE WASTIVE" invites audiences to reconsider their creative processes and consumption habits. It serves as a poetic call for more mindful, sustainable practices, provoking deeper reflections on our interconnectedness with the environment.

## AUTHORS KEYWORDS
Human-computer interaction; digital fabrication; interactive installation; sustainability

## CSS CONCEPTS
• Human-centered computing ~ Interaction Design; Applied computing ~ Media Arts; Hardware

## INTRODUCTION

Emerging technologies in HCI, such as digital fabrication, make innovative design and physical prototyping more accessible to a wider audience. However, these advancements also intensify environmental challenges, as the prototyping process consumes energy and various materials, particularly the use of wide variety of on-the-shelf plastics is increasing [9]. When improperly disposed of, these materials break down into microplastics that threaten marine ecosystems, contributing to the estimated 11–23 million tons of plastic entering the ocean each year [4].

To address this issue, Sustainable Interaction Design (SID) was introduced as a key focus [1, 2], emphasizing the integration of environmental considerations into computing. More recently, Eldy et al. proposed a sustainable prototyping life cycle for digital fabrication, offering designers guidance on minimizing waste throughout the design and prototyping process [3]. Additionally, researchers have been actively exploring eco-friendly materials [5, 6]. For example, Rivera et al. developed a sustainable 3D printing filament made from spent coffee grounds, demonstrating how organic waste can be repurposed into functional fabrication materials [5]. These initiatives reflect a growing commitment to circular design strategies that embed sustainability within digital fabrication practices.

While these are essential steps forward, they largely remain within expert communities. How can we foster broader engagement with sustainability while also exploring the creative potential of emerging technologies? What if discarded materials could observe the world? What would they see? What would they say?

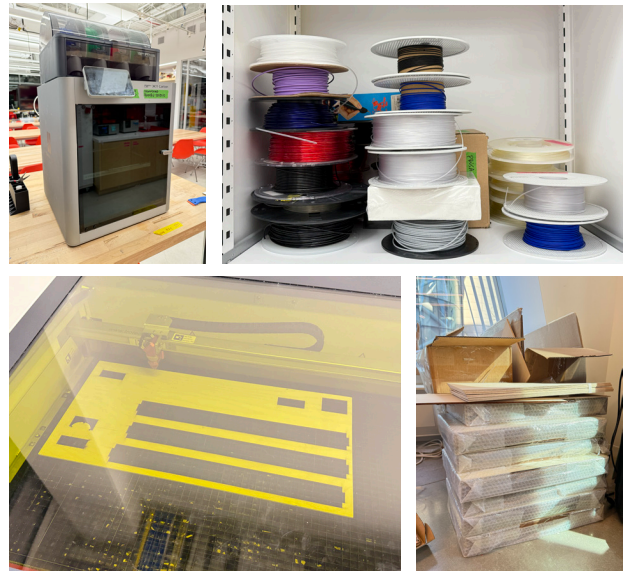

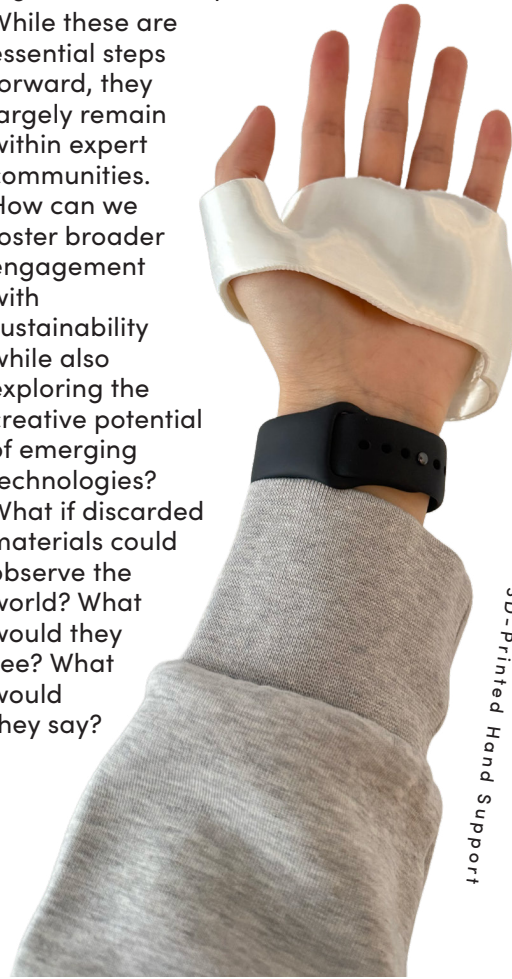

*3D-Printed Hand Support*

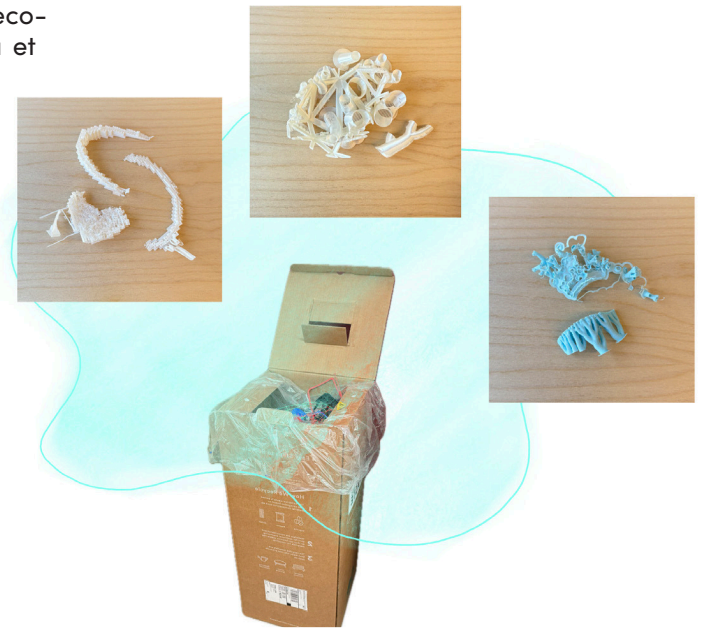

Common 3D Printing Waste: Removed support structures, excess infill, and failed prints, etc.

Typical Laser Cut Waste: plywood scraps.

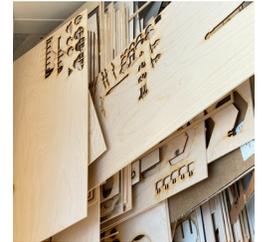

"THE WASTIVE" is an artistic response to these questions, transforming discarded digital fabrication materials into interactive, sentient observers. Rather than treating waste as inert and disposable, the installation reimagines these remnants as active participants in a silent dialogue. As viewers approach, the installation awakens, mirroring the rhythmic ebb and flow of ocean waves — a silent dialogue where discarded materials "observe" and respond to human presence. These interactions evoke the gentle murmurs of the sea, transforming technological residue into a reflective, sensory experience.

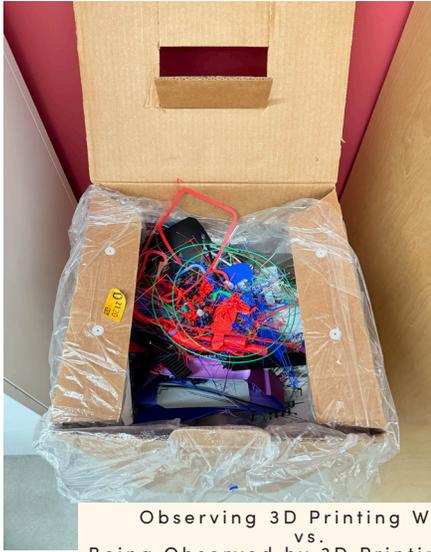

Observing 3D Printing Waste
vs.
Being Observed by 3D Printing Waste

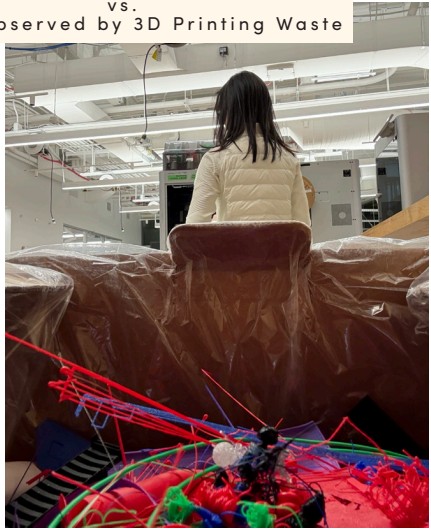

## INSPIRATION

I (the first author of this work) embarked on my digital fabrication journey with excitement during a design course, where 3D printing and laser cutting opened new creative possibilities, making prototyping more precise and efficient.

Yet, this excitement gradually gave way to an unsettling realization. One day, as I stood before a recycling bin, scraping off a 3D-printed tree structure, I was struck by the sheer volume of discarded fabrication waste — failed prints, misaligned cuts, and leftover materials piling up, often unnoticed. While these scraps were placed in recycling bins, I began to question: Were they truly being repurposed, or were they simply becoming another layer of environmental waste?

"SHADED SEAS" by Studio Swine allows audiences to manipulate digital representations of ocean plastics in real time, serving as a metaphor for human influence on environmental change [7]. On the other hand, the kinetic installation "Shylight" explores how an inanimate object can mimic changes that express character and emotions [8]. These works inspired me to consider: What if discarded fabrication waste could "observe" its surroundings, reacting to human activity rather than remaining passive?

The concept also draws from the observer effect, a principle in physics suggesting that observation alters the observed phenomenon. What if waste, seemingly lifeless, could react to human presence? In "THE WASTIVE", digital fabrication waste "observes" through a laptop camera, becoming another "occupier" that turns the audience into the observed. The audience's presence activates the discarded materials, setting the installation in motion and creating a dynamic interplay between human interaction, technological remnants, and the environment they inevitably affect.

"THE WASTIVE" reimagines digital fabrication waste not as inert, discarded remnants but as active participants in a reflective dialogue on sustainability. By transforming waste into an interactive medium, the installation invites audiences to reconsider material consumption and explore the intersection of human creativity, technology, and environmental responsibility. By merging digital fabrication waste with interactive elements, "THE WASTIVE" transforms discarded materials into a medium for storytelling, challenging audiences to rethink sustainability, recognize their role in material lifecycles, and engage in a poetic dialogue on the delicate balance between technological progress and ecological responsibility.

## CREATION AND FABRICATION

This installation utilizes digital fabrication waste generated from everyday prototyping activities on campus. While PLA (polylactic acid), a widely used material in 3D printing, is marketed as biodegradable, it only decomposes within 90 days under industrial composting conditions but can persist in landfills for 100 to 1,000 years [9]. Meanwhile, plywood, often considered a more sustainable alternative, still demands significant water resources during production [9]. By incorporating these two materials, "THE WASTIVE" emphasizes the significant environmental impact of commonly used digital fabrication materials.

The structure of the installation draws on established design techniques for these materials to create a dynamic, wave-like motion. The plywood sheet is recut into previously laser-cut shapes and patterns, incorporating a living hinge design— a commonly used laser-cutting technique that adds flexibility, enhancing the organic flow of the wave and reinforcing its sense of movement and transformation. PLA waste, gathered from various sources, is integrated into these cut forms,

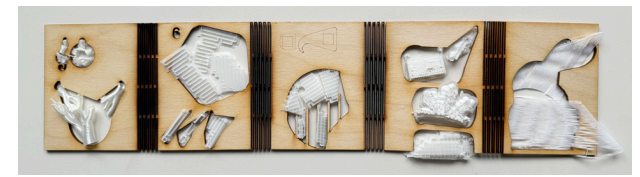

The Kinetic Panel made with Digital Fabrication Waste.

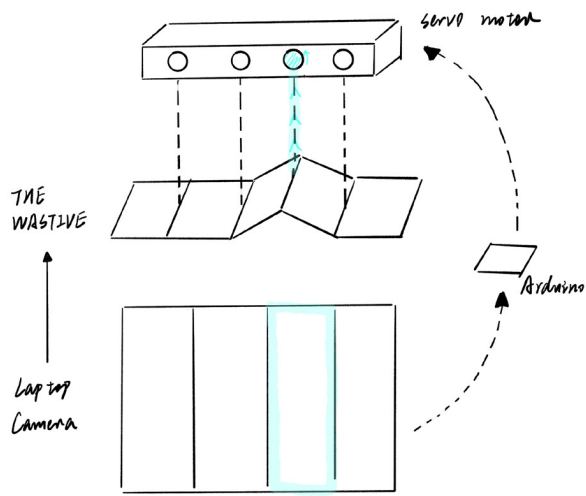

Figure 1. Schematic Structure of "THE WASTIVE".

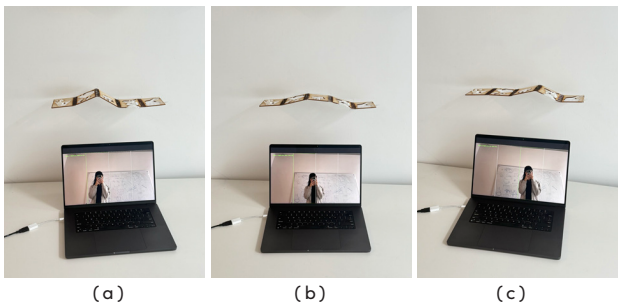

Figure 2. Dynamic Interaction between "THE WASTIVE" and Audience.

## CONCLUSION
By conveying sustainability through the medium of interactive art, "THE WASTIVE" fosters a deeper awareness of our interconnectedness with the materials we create — and discard. It serves as both a critique and an invitation: a critique of wasteful design practices and an invitation to embrace more mindful, sustainable approaches to innovation. In doing so, it opens new possibilities for human-computer interaction, where even discarded remnants become storytellers, provoking reflection on the delicate balance between technological advancement and ecological responsibility.

highlighting the intertwined nature of creation and waste.

Figure 1. depicts the responsive system at the core of "THE WASTIVE" that brings the installation to life. A laptop camera tracks the audience's position in real-time, transmitting spatial data to an Arduino microcontroller, which then activates servo motors embedded in the structure. As viewers move, the motors adjust the elevation of different sections, causing the waves to mimic the ebb and flow of the ocean.

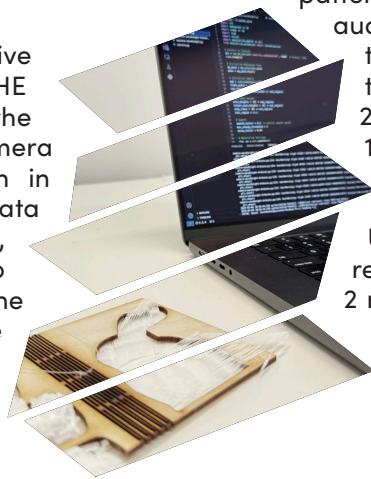

## INTERACTING WITH "THE WASTIVE"
As the audience stands in front of the installation and observes, it responds by generating waves that correspond to their position. The longer they remain in a particular area, the higher the waves rise.

Figure 2. depicts how "THE WASTIVE" responds to the audience's movement between detected regions. As they shift to a new region, the wave pattern dynamically adapts. Figure 2. (a) The audience is in region 1 (second region from the left on the laptop screen), causing the corresponding wave to rise. Figure 2. (b) As the audience moves from region 1 to region 2 (third region from the left on the laptop screen), the wave pattern transitions dynamically. Figure 2. (c) Upon reaching region 2, the wave in region 1 lowers while the wave in region 2 rises in response to the audience's new position.